\documentclass[aps,preprint]{revtex4}
\usepackage{amssymb}

\input{tcilatex}

\begin{document}

\preprint{}
\title[SIR and dynamical Monte Carlo]{\textbf{Dynamical Monte Carlo method
for stochastic epidemic models }}
\author{O.E. Ai\'{e}llo and M.A.A. da Silva}
\affiliation{Departamento de F\'{\i}sica e Qu\'{\i}mica, FCFRP,Universidade de S\~{a}o
Paulo, 14040-901 Ribeir\~{a}o Preto,SP, - Brasil.}
\keywords{Epidemic, Stochastic, Dynamical, Monte Carlo, SIR}
\pacs{02.70.Tt,05.10.Ln,02.50.Ga,87.23.-n}

\begin{abstract}
In this work we introduce a new approach to Dynamical Monte Carlo methods to
simulate markovian processes. We apply this approach to formulate and study
an epidemic generalized SIRS model. The results are in excellent agreement
with the fourth order Runge-Kutta method in a region of deterministic
solution.\ Introducing local stochastic interactions, the Runge-Kutta method
is no longer applicable. Thus, we solve the system described by a set of
stochastic differential equations by a Dynamical Monte Carlo technique and
check the solutions self-consistently with a stochastic version of the Euler
method. We also analyzed the results under the herd-immunity concept.
\end{abstract}

\volumeyear{}
\volumenumber{}
\issuenumber{}
\eid{}
\date{August 19, 2002}
\received[Received text]{date}
\revised[Revised text]{date}
\accepted[Accepted text]{date}
\published[Published text]{date}
\startpage{01}
\endpage{}
\maketitle

\section{Introduction}

Epidemic systems have been systematically and mathematically formulated in a
continuous-deterministic approach, taking immediate advantages of many
numerical methods and techniques developed to solve differential equations.
The stochastic framework is more complex to analyze because of the required
detail; therefore it could be traditionally less preferable than the
deterministic ones, even being more realistic in principle \cite{Mollison,
Murray, Yakowitz}. However, improved machine technology has spread the use
of computationally intensive methods to solve a great diversity of epidemic
models, and simulation techniques, as Monte Carlo ($MC$) \cite{Metropolis,
Binder}, are becoming more popular in this matter. Some $MC$ studies hide
the effective role of the time, on time-dependent phenomena, reporting its
results as function of integral Monte Carlo steps ($MCS$) \cite{Haas}.
Ambiguities of the relationship between $MC$ time and real time preclude
rigorous comparison of simulated results between theory and experiment.
However, in the past few years, the idea of use MC methods to simulate
dynamical processes has advanced in many publications\cite%
{Gillespie,Fichtorn,Cao,Prados}.

The aim of the present work is to present a Dynamical Monte Carlo ($DMC$)
method for simulation of markovian processes. Another purpose is to
incorporate explicit spatial components into epidemic models and analyze the
dynamics of infections spread based on this method. We will apply the method
to the compartmental Susceptible-Infected-Recovered ($SIR$) model. By the
inclusion of a reflux of susceptible into the system we obtained a variant
model: $SIRS$; i.e., once recovered the individual can turns back again to
the class of susceptibles. Mean field and local interactions will be
considered. We compare the results obtained by DMC, for mean field models,
with Runge-Kutta method. In cases where Runge-Kutta method is not
applicable, the $MC$ space-dependent results are checked self-consistently
using a stochastic version of the Euler method \cite{Aielo} and analyzed
under the herd-immunity concept \cite{Thomas}.

We subdivided the work in the following way: method description (section
II), Monte Carlo Simulation technique (section III), \ epidemic models
(section IV), and finally we apply the methodology for the solution of the $%
SIRS$ model (section V).

\section{The method}

Stochastic process approaches could simulate non-equilibrium systems, even
the deterministic ones, introducing random variables to describe them in a
microscopic scale. The macroscopic behavior of some system is resulting from
averages of its microscopic properties. Here, we will simulate systems only
with markovian processes. Thus, we describe the evolution of the
distribution of probabilities with the \emph{Master Equation}: 
\begin{equation}
\frac{dP_{i}(t)}{dt}=\sum_{j}w_{j\rightarrow
i}P_{j}-\sum\limits_{j}w_{i\rightarrow j}P_{i},
\label{Pauli Master Equation}
\end{equation}%
where $P_{i}(t)$ is the probability to find the system in the state $i$ in
the time $t$, and $w_{i\rightarrow j}$ is the probability of transition per
unit of time. The first term on the right side describes the rate of all
transitions to reach the considered state (increasing its probability), and
the second term describes the rate of all transitions leaving the considered
state (decreasing its probability). Considering $T_{ij}$ as the probability
of transition from the state $i$ to $j$, we can write $w_{ij}=\frac{T_{ij}}{%
\tau _{i}}$ \cite{Hoel}, where $\tau _{i}$ is the characteristic time
constant (\textit{lifetime}) of the state $i$. The probabilities, $P_{i}(t)$
and $T_{ij}$, obey the normalization conditions: $\sum_{i}P_{i}(t)=1$ and $%
\sum_{j}T_{ij}=1$.

We now start by choosing a convenient physical extensive microscopic
quantity $A_{i}$, which depends only of the system's state $i$. Since the
time must change for every successful event, for our purposes here, from now
on we will consider only counting events related quantities. The mean value
for a given quantity at the time $t$ is%
\begin{equation}
A(t)=\langle A\rangle =\sum_{i}P_{i}(t)A_{i}.  \label{Temporal Mean Value}
\end{equation}%
This equation represents the macroscopic physical quantity $A(t)$.
Differentiating both sides of the equation above with respect to $t$, we
obtain 
\begin{equation}
\frac{dA(t)}{dt}=\sum_{i}\frac{dP_{i}(t)}{dt}A_{i},
\label{Time Evolution Equation}
\end{equation}%
and by substituting $\left( \ref{Pauli Master Equation}\right) $ in $\left( %
\ref{Time Evolution Equation}\right) $ follows: 
\begin{equation}
\frac{dA(t)}{dt}=\sum_{i}\sum_{j}w_{j\rightarrow
i}P_{j}A_{i}-\sum_{i}\sum_{j}w_{i\rightarrow j}P_{i}A_{i}.
\label{Macroscopic Master Equation 1}
\end{equation}%
Defining $\Delta A_{ij}=A_{i}-A_{j}$ , and as $i$ and $j$ sweep all possible
states of the system, we may rewrite $\left( \ref{Macroscopic Master
Equation 1}\right) $ as 
\begin{equation}
\frac{dA(t)}{dt}=\sum_{i}\sum_{j}w_{j\rightarrow i}P_{j}\Delta A_{ij}.
\label{Macroscopic Master  Equation 2}
\end{equation}

Consider now a discretized system with $N$ interacting elements. Each
element has $g$ degrees of freedom given by the set $\left\{ \gamma
_{i}\right\} $ with $g$ dynamic variables $\gamma _{i}$, $i=1,2,..,g$. By
doing an element \textit{move}, i.e., changing some $\gamma _{i}$ value of a
chosen element, the system will reach a next-neighbor microscopic state.
Suppose that we are in a time scale where only one event occurs, and each
event produces only one element \textit{move}. In another words, we are
neglecting transitions between \ states that need more than one element 
\textit{move} to take place. Thus, let us measure ``distances'' among the
states, say with the amount $\left| \Delta A_{ij}\right| $, with non null
minimum value $\left| \Delta A_{ij}\right| =$ $q$ \ that defines the
distance between first neighbor states. With the above considerations, an
approach to the equation $\left( \ref{Macroscopic Master Equation 2}\right) $
can be done as%
\begin{equation}
\frac{dA(t)}{dt}=\sum_{(ij)}w_{j\rightarrow i}P_{j}\text{$q$}\delta _{ij},
\label{Macroscopic Master Equation 3}
\end{equation}%
in which the symbol $(ij)$ denotes a pair of first neighbor states, and $%
\delta _{ij}=$ $\Delta A_{ij}/\left| \Delta A_{ij}\right| $. Now consider
other physical quantity $A^{\dagger }$ as the \textit{source} for the
quantity $A$. The use of the term \textit{source} here is in the following
sense: increasing $A$ by the quantity $q$, $A^{\dagger }$ decreases by the
same quantity and vice-versa. Thus, we can rewrite $\left( \ref{Macroscopic
Master Equation 3}\right) $ as:%
\begin{equation}
\frac{dA(t)}{dt}=\sum_{j}r_{j}^{+}P_{j}A_{j}^{\dagger
}-\sum_{j}r_{j}^{-}P_{j}A_{j},  \label{Macroscopic Master Equation 4}
\end{equation}%
where the rate $r_{j}=\left\langle w_{j\rightarrow i}\right\rangle _{i}$
results from the average of the transition probabilities per unit of time,
over the \textit{ensemble} of first neighbor states $i$ of $j$ in the time $%
t $, i.e., the mesoscopic rates. The word \textit{ensemble} here means a
group of accessible configurations in a small time interval around the time $%
t$. We are using the time dependent ergodicity idea\cite{Binder2}, and in
this sense, usually, the systems are non ergodic in non equilibrium states.
The superscripts ``$+$'' and ``$-$'' label the contributions to increase and
to decrease the quantity $A(t)$, respectively. In the particular cases where 
$r_{j}^{+}=r^{+}$ and $r_{j}^{-}=r^{-}$ are constant (or only function of
the time) we have:%
\begin{equation}
\frac{dA}{dt}=r^{+}A^{\dagger }-r^{-}A,  \label{Flux Master Equation}
\end{equation}%
which is similar to the kinetic equation for chemical reactions of first
order $\mathcal{A}^{\dagger }\rightleftarrows \mathcal{A}$, being $%
A^{\dagger }$ and $A$ the respective concentrations of the chemical elements 
$\mathcal{A}^{\dagger }$ and $\mathcal{A}$. The system is in equilibrium
when the balance at macroscopic level: $r^{+}A^{\dagger }=r^{-}A$, is
satisfied. We can reach the equilibrium imposing the detailed balance,
although this assumption is not necessary\cite{Binder}; as we will see, the
equilibrium is consequence from the chosen hierarchy, that solves the Master
Equation in any instant.

To solve the equation $\left( \ref{Macroscopic Master Equation 3}\right) $
we write it in the integral form:%
\begin{equation}
A(t)-A(t_{0})=\int_{t_{0}}^{t}\tsum_{(ij)}w_{j\rightarrow i}P_{j}(t)\text{$q$%
}\delta _{ij}dt.  \label{Integral Master Equation}
\end{equation}%
Discretizing the equation $\left( \ref{Integral Master Equation}\right) $,
we can write:%
\begin{equation}
A(t)-A(t_{0})\simeq \sum_{k=0}^{n}\sum_{(ij)}w_{j\rightarrow i}P_{j}(t_{k})%
\text{$q$}\delta _{ij}\Delta t_{k}.
\label{Discrete integral master equation}
\end{equation}%
Let the group of possible probabilities of transition per unit of time $%
w_{j\rightarrow i}$ represented by the set $\mathcal{P}_{t}=\{w_{j%
\rightarrow i}\}$, being the $i$ and $j$ states occurring around an instant $%
t$, with $w_{t}^{\max }=$ $\sup \mathcal{P}_{t}$, that is, the largest
probability in $\mathcal{P}_{t}$. Each element of the time in the equation $%
\left( \ref{Discrete integral master equation}\right) $ could be%
\begin{equation}
\Delta t_{k}=\frac{1}{w_{t_{k}}^{\max }N}.  \label{Delta T Fixed}
\end{equation}%
Starting from some initial condition, we can do the following iterative
process:%
\begin{equation}
A(t_{k+1})=A(t_{k})+\sum_{(ij)}w_{j\rightarrow i}P_{j}(t_{k})\text{$q$}%
\delta _{ij}\Delta t_{k}.  \label{Euler}
\end{equation}%
At each step $k$ a time interval $\Delta t_{k}$ is calculated using $\left( %
\ref{Delta T Fixed}\right) $. The probabilities per unit of time $%
w_{j\rightarrow i}\in \mathcal{P}_{t_{k}}$ are randomly drawn using the
hierarchy described in the next section of this work. Repeating the
procedure, in a sufficient number, to get a good sample of $A(t_{k})$, we
estimate the averages of $A(t_{k})$ for every $t_{k}$. This procedure is a
stochastic version of the Euler method.

\section{Monte Carlo Simulation}

In a dynamical interpretation, the $MC$ method provides a numerical solution
to the \emph{Master Equation}. In order to do this, a sequence of events is
generated based on the transition probabilities. The task of the $MC$
algorithm is to create a chronological sequence of the distinct events
separated by certain interevent times. In according to the hypothesis that
leads to the equation $\left( \ref{Macroscopic Master Equation 3}\right) $,
these interevent times are on a scale at which no two events occur
simultaneously.

To find a hierarchy to the $MC$ algorithm, we consider $n=lN$, with $l$
sweepings on the discretized system phase space, in which we are measuring a
physical quantity represented in the equation $\left( \ref{Discrete integral
master equation}\right) $; in the limit of $N\rightarrow \infty $ we have
the exact solution of the integral $\left( \ref{Integral Master Equation}%
\right) $ for a given initial condition. With this consideration, and using
the expression $\left( \ref{Delta T Fixed}\right) $, the equation $\left( %
\ref{Discrete integral master equation}\right) $ goes to the form:%
\begin{equation}
A(t)-A(t_{0})=\sum_{k=0}^{\ell N}\sum_{(ij)}\left( \frac{w_{j\rightarrow i}}{%
w_{t_{k}}^{\max }}\right) \left( \frac{1}{N}\right) P_{j}(t_{k})\text{$q$}%
\delta _{ij}.  \label{Discrete integral 2}
\end{equation}%
We can, thus, create a hierarchical process choosing the transition
probabilities as: 
\begin{equation}
T_{j\rightarrow i}^{\ast }=\frac{w_{j\rightarrow i}}{w_{t_{k}}^{\max }},
\label{Transition probabilities}
\end{equation}%
which reproduces the correct frequencies of events in every time $t_{k}$ to
solve $\left( \ref{Discrete integral 2}\right) $.

To execute the $MC$ procedure, an element is randomly selected with
probability $1/N$, and thus an attempted move, with probability given by $%
\left( \ref{Transition probabilities}\right) $, is done to change the state
from $j$ to $i$. Therefore, an event that changes a dynamic variable $\gamma
_{i}$, of the chosen element, controls the microscopic transition $%
j\rightarrow i$. When the system has \ ``degeneracy'' \ as for the events
occurrence, we need to decide what event will have chance (given by $\left( %
\ref{Transition probabilities}\right) $) to take place; thus, we chose one
of them with equal \textit{a priori} probability\cite{Toda}, supposing a 
\textit{local equilibrium} over the time. The \textit{local equilibrium}
hypothesis means that we can measure the properties of the system at any
instant $t$. Repeating the procedure, the space is swept $\mathit{l}$ times,
with the increment of time in each $MCS$\ given by $\left( \ref{Delta T
Fixed}\right) $, up to the system reach some desired final time. We denoted
1 $MCS$ as a single trial to change the state of the system. Beginning with
the same initial condition for the physical quantities, repeating the whole
process described above, we obtain the average quantity $A(t)$ over each
instant $t$. As a given state is chosen with its correct probability in a
given time, an ideal $MC$ procedure leads to%
\begin{equation}
A(t)-A(t_{0})=\sum_{k=0}^{\ell N}\left( \left\langle r^{+}A^{\dagger
}\right\rangle _{j_{k}}-\left\langle r^{-}A\right\rangle _{j_{k}}\right)
\left( \frac{1}{w_{t_{k}}^{\max }N}\right) ,  \label{Discrete Integral 3}
\end{equation}%
where the averages are done over the \textit{ensemble} of the $j_{k}$ states
in each instant $t_{k}$.

Observing some points is necessary: first, generally different runs give
different time results $t_{k}$ at the same $MCS$ $k$, and we obtain the
sample average with either linear interpolation or extrapolation data group,
in each $MC$ realization of the system\cite{footnote}, as we will describe
below. Second, in a complete sweep around a time $t_{k}$, the value $%
w_{t_{k}}^{max}$ should be approximately constant in order not to change the
hierarchy and consequently the result. Third, as the configurations do not
vary drastically in few steps, the microscopic transitions reproduce the
mesoscopic results.

Another approach to calculate the real time consists in estimating the
interevent times with the following rule: 
\begin{equation}
\Delta t_{k}^{e}=\frac{f_{e}^{k}\text{$q$}}{r_{j_{k}}^{e}A_{j_{k}}^{e}},
\label{Interevent time}
\end{equation}%
where $r_{j_{k}}^{e}=r_{jk}^{+}$ and $A_{j_{k}}^{e}=A_{jk}^{\dagger }$, or $%
r_{jk}^{e}=r_{j_{k}}^{-}$ and $A_{j_{k}}^{e}=A_{j_{k}}$ depending, whether
the result of the experiment increases or decreases the quantity $A$. The
quantity $f_{e}^{k}$ is an $e$-event factor dependent and it obeys the
relationship $\sum_{e}f_{e}^{k}=1$ (normalization condition), for each time $%
t_{k}$. We note that the time given by $\left( \ref{Interevent time}\right) $
represents the mean waiting time for transitions from a given state $j_{k}$
to any neighbor state $i$; if the microscopic state stays unchanged, the
time also does not change. We can show that this procedure leads to the same
result found using $\left( \ref{Delta T Fixed}\right) $ in each $MCS$,
observing that%
\begin{equation}
\Delta t_{k}=\sum_{e}\sum_{i}\left( \frac{w_{j_{k}\rightarrow i}}{%
w_{t_{k}}^{\max }}\right) \left( \frac{1}{N}\right) \Delta t_{k}^{e}.
\label{Delta T fixed 2}
\end{equation}%
Using the equation $\left( \ref{Interevent time}\right) $, the normalization
condition for $f_{e}^{k}$ and the definition $r_{j_{k}}^{e}A_{j_{k}}^{e}=q%
\sum_{i}w_{j_{k}\rightarrow i}$ in $\left( \ref{Delta T fixed 2}\right) $,
we obtain the expression $\left( \ref{Delta T Fixed}\right) $. In
particular, if we choose $f_{e}^{k}=0$, for every event, except $e=s$, we
have $f_{s}^{k}=1$. Under this condition, the time between $s$-events is the
waiting time. Based on this, to estimate the waiting time in a
coarse-grained way, we may define $f_{e}^{k}\equiv n_{e}^{k}/\mathcal{N}_{k}$%
;\ where $n_{e}^{k}$ is the number of $e-$events, and $\mathcal{N}%
_{k}=\sum_{e}n_{e}^{k}$ is the total number of events, in a time interval
(arbitrary) near to some time $t_{k}$.

Note that at each MCS, the minimum quantity $q$ is either added or
subtracted from the resulting quantity $A$ following the prescribed
hierarchy. This procedure, in according to $\left( \ref{Discrete Integral 3}%
\right) $ reproduces statistically the average quantity $A(t)$. Therefore,
since we have the rates, or the probabilities of transition per unit of
time, defining the time intervals between events in some scale, we construct
a $MC$ algorithm to solve the \emph{Master Equation}; consequently, we
obtain the time evolution of physical quantities of the system.

In order to define the errors on macroscopic quantities, we will do a direct
approach to calculate average quantities. We start supposing a \textit{local
equilibrium} of the system over some instant $t_{0}$. With use of
appropriated transition rates, we can reach any state $i$ with probability $%
P_{i}(t_{0})$; so constructing several independent markov chains generates
the distribution, which produces an \textit{ensemble} of configurations over
the time $t_{0}$. Thus, we can use directly \ the equation $\left( \ref%
{Temporal Mean Value}\right) $ to calculate $A(t)$ at the time $t_{0}$. If
we chose a given state $i$ of the system with probability $P_{i}^{\ast
}(t_{0})$, we may rewrite the equation $\left( \ref{Temporal Mean Value}%
\right) $ by\cite{Binder} 
\begin{equation}
A(t_{0})=\frac{\sum_{i}P_{i}(t_{0})A_{i}/P_{i}^{\ast }(t_{0})}{%
\sum_{i}P_{i}(t_{0})/P_{i}^{\ast }(t_{0})}.  \label{Ensemble Average}
\end{equation}%
A natural choice of $P_{i}^{\ast }(t_{0})$, under equilibrium, is $%
P_{i}^{\ast }(t_{0})=P_{i}(t_{0})$, obtaining 
\begin{equation}
A(t_{0})=\frac{\sum_{i=1}^{L^{\ast }}A_{i}}{L^{\ast }},  \label{MC average}
\end{equation}%
where $L^{\ast }$ is the number of all possible states of the system at the
time $t_{0}$. This result extends readily to any time $t$. The states
labeled by $i$ may be considered as virtual states corresponding to possible
data interpolation or extrapolation. The practical procedure is the
following: at a given $MC$ realization of the system (experiment), in the
construction of a trajectory, labeled by $\ell $, we may get the
measurements of any appropriated physical quantity $A_{\ell }(t)$ obtained
by either linear extrapolation or interpolation using two consecutive data
points. After perform $L$ Monte Carlo experiments, at some time $t$, the
mean value of $A$ is $A(t)\approx \sum_{\ell =1}^{L}\frac{A_{\ell }}{L}$.
Note that if we idealize this procedure doing $L\rightarrow \infty $, we
obtain the complete \textit{ensemble} that give-us the correct mean values
of physical quantities for each time $t$. Ensuring that different
experiments are independent, the error for the involved quantities in the
process for each time $t$ could be\cite{Binder}: 
\begin{equation}
\frac{\sigma _{A}}{\sqrt{L}}=\sqrt{\frac{<A^{2}>-<A>^{2}}{L}},
\label{Error Equation}
\end{equation}%
where $A$ can be, for example, in this work context, the number of infected
individuals.

\section{Epidemic models}

The conventional treatment of epidemic systems is formulated based on a
group of compartments that represents each of the possible \emph{statuses},
of its elements, for which we may assign dynamic variable values, with rates
of transfer among pairs of compartments. Mathematically this subject turns
into a set of differential equations. Considering a generic system
(population, epidemic agents, etc.) and its space distributions, the
temporal and space evolution characterizes any epidemic, and in each region,
the density of the elements can vary with the time. Under this optics we
considered the epidemic phenomenon as a stochastic process in which one
random variable is the time. This focus seeks to propitiate the
incorporation of more details in the study of epidemic process and to allow
the analysis of more complex models and therefore more realistic.

The $SIRS$ model considers a population with $N$ individuals divided in
three classes: \textbf{S} (susceptible individuals), \textbf{I} (infected)
and \textbf{R} (recovered). The evolution of the disease occurs according to
the outline \textbf{S}$\rightarrow $\textbf{I}$\rightarrow $\textbf{R}$%
\rightarrow $\textbf{S}. Based on the equation $\left( \ref{Macroscopic
Master Equation 4}\right) $, we formalize the $SIRS$ model in a quite
generic way through the following group of differential equations:%
\begin{eqnarray}
\frac{dS}{dt} &=&\ \sum_{j}r_{j}^{\text{\textbf{RS}}}P_{j}R_{j}-%
\sum_{j}r_{j}^{\text{\textbf{SI}}}P_{j}S_{j},  \label{DSDT} \\
\frac{dI}{dt} &=&\sum_{j}r_{j}^{\text{\textbf{SI}}}P_{j}S_{j}-\sum_{j}r_{j}^{%
\text{\textbf{IR}}}P_{j}I_{j},  \label{DIDT} \\
\frac{dR}{dt} &=&\sum_{j}r_{j}^{\text{\textbf{IR}}}P_{j}I_{j}-\
\sum_{j}r_{j}^{\text{\textbf{RS}}}P_{j}R_{j},  \label{DRDT}
\end{eqnarray}%
where $S$, $I$ and $R$ are the (average) number of susceptible, infected and
recovered individuals, respectively, over each instant $t$. The mesoscopic
rates are $r_{j}^{\text{\textbf{SI}}}$, $r_{j}^{\text{\textbf{IR}}}$ and $%
r_{j}^{\text{\textbf{RS}}}$, for each state $j$, from \ \textbf{S}$%
\rightarrow $\textbf{I} , \textbf{I}$\rightarrow $\textbf{R}\ and \textbf{R}$%
\rightarrow $\textbf{S}, respectively.

In order to reproduce the deterministic model in the reference \cite{Aielo2}
we did the following restrictions:

1) the effective increase rate of those susceptible (individuals) is
directly proportional to the number of recovered, $\sum_{j}r_{j}^{\text{%
\textbf{RS}}}P_{j}R_{j}=mR$; and consequently the recovered decrease in the
same proportion;

2) the effective increase rate of those infected is directly proportional to
the number of infected and a power $\mu $ of susceptible, $\sum_{j}r_{j}^{%
\text{\textbf{SI}}}P_{j}S_{j}=bS^{\mu }I/N^{\mu }$; and consequently the
susceptible decrease in the same proportion;

3) the effective removal rate of those infected is directly proportional to
the number of infected, $\sum_{j}r_{j}^{\text{\textbf{IR}}}P_{j}I_{j}=aI$;
and consequently the recovered increase in the same proportion. Taken these
restrictions to the set of differential equations $\left( \ref{DSDT}-\ref%
{DRDT}\right) $ give:%
\begin{eqnarray}
\frac{dS}{dt} &=&mR-\frac{bS^{\mu }I}{N^{\mu }}  \label{SIRS1} \\
\frac{dI}{dt} &=&\frac{bS^{\mu }I}{N^{\mu }}-aI\   \label{SIRS2} \\
\frac{dR}{dt} &=&aI-mR,  \label{SIRS3}
\end{eqnarray}%
in which $\mu $ relates to the \textit{safety-in-numbers power}\cite{Bailey}%
. The conditions: $dS/dt=0$, $dI/dt=0$, $dR/dt=0$, determines the
steady-state solutions; the nontrivial solution occurs for finite values of $%
\ S_{\sigma }$, $I_{\sigma }$, and $R_{\sigma }$, viz.,%
\begin{eqnarray}
S_{\sigma } &=&(a/b)^{1/\mu }N,  \label{Equilibrium S} \\
I_{\sigma } &=&\frac{1-(a/b)^{1/\mu }}{1+a/m}N,  \label{Equilibrium I} \\
R_{\sigma } &=&\frac{1-(a/b)^{1/\mu }}{1+m/a}N.  \label{Equilibrium R}
\end{eqnarray}%
Depending on the removal rate $a$ of the infectives, infection parameter $b$%
, and renewal $m$, there exist stable solutions around the steady state that
correspond to recurrent epidemics, or damped (fading) recurrent waves. These
variant supplies oscillatory solutions that vanish with the time, reaching a
stationary state, in which, the number of elements in each class stays
constant. This model is a generalization of the classical $SIR$ system \cite%
{Murray,Kermack,Weiss}, readily recovered from $\left( \ref{SIRS1}-\ref%
{SIRS3}\right) $ by setting $\mu =1$ , $m=0$, that gives\cite{Anderson79}:%
\begin{eqnarray}
\frac{dS}{dt} &=&-bSI,  \label{SIR1} \\
\frac{dI}{dt} &=&bSI-aI,  \label{SIR2} \\
\frac{dR}{dt} &=&aI.  \label{SIR3}
\end{eqnarray}%
The $SIR$ class of compartmental models has several deterministic and
stochastic versions, as the $SIS$ and the $SEIR$ model \cite%
{Weiss,Rhodes,Johansen}. With no inclusion of spatial variables, they are
often considered as deterministic mean field models, based on the chemical ``%
\textit{mass action}'' principle.

In this work, we considered epidemic processes as a result from the action
of a mean field and the interaction among the closest individuals (local
interaction). To promote the infection by the contact between infected
individuals and susceptibles, a stochastic term is added to the
deterministic $SIRS$ model. Therefore, the transition probabilities per unit
of time became%
\begin{eqnarray}
w_{\text{\textbf{R}}\rightarrow \text{\textbf{S}}} &=&m,  \label{WRS} \\
w_{\text{\textbf{S}}\rightarrow \text{\textbf{I}}} &=&\Gamma \,\frac{b}{%
N^{\mu }}S^{\mu -1}I+\Lambda \,[1-(1-p_{0})^{n}],  \label{WSI} \\
w_{\text{\textbf{I}}\rightarrow \text{\textbf{R}}} &=&a.  \label{WIR}
\end{eqnarray}%
The $\Gamma $ and $\Lambda $ parameters balance, the global (mean field) and
the local (nearest neighbors) variables, respectively; the relation $\Gamma
+\Lambda =1$ is satisfied. The parameter $p_{0}$ is the probability for a
susceptible to become infected due to a unique infected neighbor. Therefore, 
$(1-p_{0})^{n}$ is the probability of no infection of a susceptible if it
has $n$ infected neighbors, thus $1-(1-p_{0})^{n}$ is the probability of
infection of a susceptible if it has $n$ infected neighbors. The standard
infection rate $b,$ recovery rate $a,$ exponent $\mu $ and the renewal rate$%
\ m$ are positive $\mathcal{O}(1)$ parameters, and $S(t)+I(t)+R(t)=N,$ with $%
dN/dt=0.$ When the renewal parameter is non zero ($m\neq 0)$, a continuous
influx of susceptible rises up into the system, producing oscillations in
the number of elements of the populational class $\mathcal{C=}\{S,I,\,R\}$.
Therefore, fading recurrent epidemics may occur before it reaches the steady
(endemic) state. The power $\mu $ introduces a modification in the original $%
SIR$ model that takes in account nonhomogeneous mixing of susceptible and
infective.

When only the mean field interaction is considered, the Runge-Kutta method
is enough to solve the $SIRS$ model. However, the $DMC$ method, besides to
solve systems with local interactions, also supplies the stochastic dynamic
one.

\section{Application of the $DMC$ to the $SIRS$ model}

In this work we consider a square lattice of $\ N=M\times M$ sites with $%
M=200$. The initial condition for the system is set up by randomly
distributing $I_{0}$ infectives on the lattice ($N>>I_{0}$) and the
remaining sites occupied by $S_{0}=N-I_{0}$ susceptibles; therefore, $%
R_{0}=0 $. The simulation develops systematically by choosing one site of
the lattice at a random at a time. Depending on its \emph{status}
(susceptible, infected or recovered), a trial to go to another \emph{status}
is done through a set of transition probabilities given by $\left( \ref{WRS}-%
\ref{WIR}\right) $, properly updating the populational class $\mathcal{C}$.
If the transition is successful, the system is now in a new state, and so we
assign a time delay to this transition. We repeat the process until the
system reaches the steady state.$\ $

In order to construct a hierarchical process, we set the probability of
transition $T_{k,\alpha }^{\ast }$, at the $MCS$ $k$, for a particular event 
$\alpha $ (\textbf{S}$\rightarrow $\textbf{I}$,\,$\ \textbf{I}$\rightarrow $%
\textbf{R}$,$ or \textbf{R}$\rightarrow $\textbf{S}$,$ in our case), in
accord with $\left( \ref{Transition probabilities}\right) $, as follows:%
\begin{equation}
T_{k,\alpha }^{\ast }=w_{\alpha }/w_{\max },
\end{equation}%
where $w_{\alpha }\in \mathcal{P}=\,\{w_{\text{\textbf{S}}\rightarrow \text{%
\textbf{I}}},w_{\text{\textbf{I}}\rightarrow \text{\textbf{R}}},w_{\text{%
\textbf{R}}\rightarrow \text{\textbf{S}}}\}$ is the transition probability
per unit of time for the event $\alpha $, and $w_{\max }=\sup \mathcal{P}$.
Thus, each particular trial is gauged according to a balance of rates,
producing a hierarchical sequence of events. \ Operationally, we compare $%
T_{k,\alpha }^{\ast }$ against a random number, $0\leq \mathcal{R}_{1}\leq 1$%
, taken from a uniform distribution. When $\mathcal{R}_{1}$ $>$\ $%
T_{k,\alpha }^{\ast }$, we reject the new state; otherwise accept it and
calculate an incremental random time $\Delta t_{k}^{\alpha }$ from $\left( %
\ref{Interevent time}\right) $, with $q${\large {\Large \ }}$=1$, as
follows: $\Delta t_{k}^{\text{\textbf{RS}}}$ $=\frac{f_{S_{+}}^{k}}{r_{k}^{%
\text{\textbf{RS}}}S^{\dagger }}$, or, $\Delta t_{k}^{\text{\textbf{SI}}}$ $=%
\frac{f_{I_{+}}^{k}}{r_{k}^{\text{\textbf{SI}}}I^{\dagger }}$, or, $\Delta
t_{k}^{\text{\textbf{IR}}}$ $=\frac{f_{I_{-}}^{k}}{r_{k}^{\text{\textbf{IR}}%
}I}$, where $f_{I_{+}}^{k}=\frac{n_{I_{+}}^{k}}{\mathcal{N}_{k}}$, \ $%
f_{I_{-}}^{k}=\frac{n_{I_{-}}^{k}}{\mathcal{N}_{k}}$, $\ f_{S_{+}}^{k}=\frac{%
n_{S_{+}}^{k}}{\mathcal{N}_{k}}$, $I^{\dagger }=S$, and $S^{\dagger }=R.$
The numbers: $n_{I_{+}}^{k}$, $n_{I_{-}}^{k}$, and $n_{S_{+}}^{k}$, are
numbers of events that increase $I$, decrease $I$ and increase $S$,
respectively, at the time $t_{k}$. The number $\mathcal{N}%
_{k}=\sum_{e}n_{e}^{k}$ is the total number of events, and the rates are: $%
r_{k}^{\text{\textbf{RS}}}=$ $w_{\text{\textbf{R}}\rightarrow \text{\textbf{S%
}}}=m$, $r_{k}^{\text{\textbf{SI}}}=\Gamma \,\frac{b}{N^{\mu }}S^{\mu
-1}I+\Lambda <\,[1-(1-p_{0})^{n}]>_{j_{k}}$, $r_{k}^{\text{\textbf{IR}}}=w_{%
\text{\textbf{I}}\rightarrow \text{\textbf{R}}}=a$. The average $<>_{j_{k}}$%
was estimated doing the sum of all local interaction terms over the system
configuration $j_{k}$, sweeping only the susceptible individuals. Overall
sum and search of $w_{\max }$, was in fact done only once, at the beginning
of the simulation, after that we updated tables. We accumulated the number
of events to obtain the factors, $\ f_{\alpha }^{k}$, in two ways: first,
over each 100 $MCS$; thus, each new time interval was determinated using the
former calculated factor, except in the first 100 $MCS$, in which we
progressively calculated the factors. Second way, we let the number of
events progressively increase, and thus, calculated the factors at each
step. As the results agreed for both approaches, we adopted the second one
because the averages converged faster.

The figures, 1 and 2, show the temporal evolution of $S(t)$, $I(t)$ and $%
R(t) $ for $SIR$ and $SIRS$ models respectively. Continuous lines represent
numerical (fourth-order Runge-Kutta) \emph{checking solutions}, and open
circles correspond to the $DMC$ simulation. Accuracies of the numerical
solutions were checked using the steady state exact solution and the
estimate of the errors were less than $0.1\%$. The results, shown in these
figures, with respect to the $DMC$ simulation correspond to an average of $%
20 $ independent trajectories, a number sufficient to produce soft curves
and illustrate the agreement with the \emph{checking solutions. }

We introduce now the local term, with the weight $\Lambda $, and the
variable $n$ \ as an integer in the interval from $n=0$ up to $8,$ since
first and second nearest infected neighbors are indistinguishably considered
for each susceptible. From a computational point of view, the main
consequence of introducing space-dependent variables is that the Runge-Kutta
method is no longer applicable to the resulting model. To check the
self-consistency of the approach we integrate numerically $\left( \ref{DSDT}-%
\ref{DRDT}\right) $, using the Stochastic Euler method described in Section
II; we calculated the $S,$ $I$ and $R$ quantities with iterations and chose
the rates randomly as in the $MC$ procedure. $MC$ solutions were checked
with this method, showing excellent agreement\cite{Aielo}, less then $1\%$
of difference, results not shown. The time evolution (number of infective)
shown in Figure 3 \emph{\ }correspond to $\Lambda =0.1$, $0.5$ and $0.9$,
and $p_{0}=0.1$. Note that increasing $\Lambda $ \ the epidemic severity
reduces. Therefore, those epidemic outbreak mechanisms involving only local
contacts are less efficient than those whose propagation is due to some
wider-range mechanisms. Note also, that for larger $\Lambda $ the second
peak of the curve displaces significantly to the right. The establishment of
a protecting shield (herd immunity effect) may explain this effect.
Depending on local contact probability, $(1-(1-p_{0})^{n})$, the size of the
removal class interferes essentially in the infection mechanism because the
number of infectives of the neighborhood (shield effect) determines the
infective character of the neighborhood of one susceptible. Figure 4
illustrates graphically the shield effect.

\section{Conclusions}

In this work we examined and applied the Dynamical Monte Carlo method to the
epidemic $SIRS$ model. We showed that, once established the hierarchy and
the relationship between Monte Carlo step and real time, we simulate the
dynamic aspects of the system, including properties out of the equilibrium.
Therefore, we can use the power and the generality of the Monte Carlo
simulation to obtain the temporal evolution of deterministic or stochastic
systems.

We emphasize that, here, are not required uncorrelated events as they were
in the reference\cite{Aielo2}. The results for independent runs need to be
uncorrelated, so we can properly use the averages obtained for each time $t$
to represent the physical quantities of the process. In order to do this we
use a local equilibrium hypothesis, what may be at first glance restrictive.
However we may even reduce the time observation enough, for the system does
not have time to leave some metastable states, say order of the lifetime $%
\tau _{i}$; we can so obtain the averaged quantities. We can obtain a good
convergence to the ideal averages, in the practice of the simulation, by
increasing the number of observations, i.e., the number of time experiments.

The system studied is sufficiently general to illustrate several aspects of
the real-time evolution determined by Dynamical Monte Carlo simulation.

The authors gratefully acknowledge funding support from FAPESP Grant n.
00/11635-7 and 97/03575-0. The authors would also like to thank Drs. A.
Caliri and M. C. Nonato for many stimulating discussions and suggestions.

\newpage

\begin{center}
{\Large Figure Captions}
\end{center}

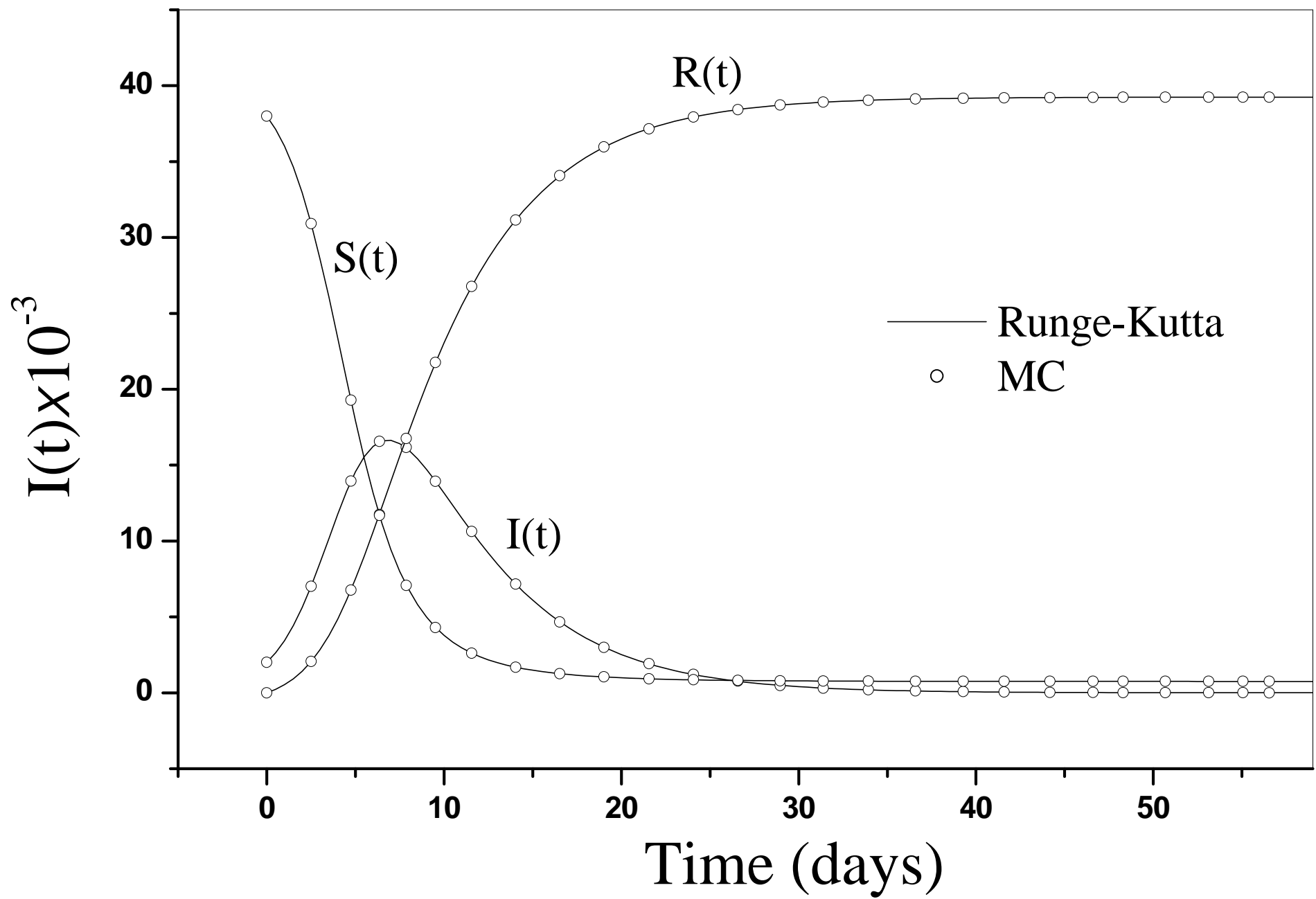
FIG. 1. $SIR$ model. The figure shows the evolution of the number of
susceptible $S$, infective $I$ and recovered $R$ with time $t$. The
numerical values for the model parameters are $r=0.2$, $b=0.8$. There is a
good agreement between the $MC$ results (open circles) and solutions
provided by Runge-Kutta method (line)

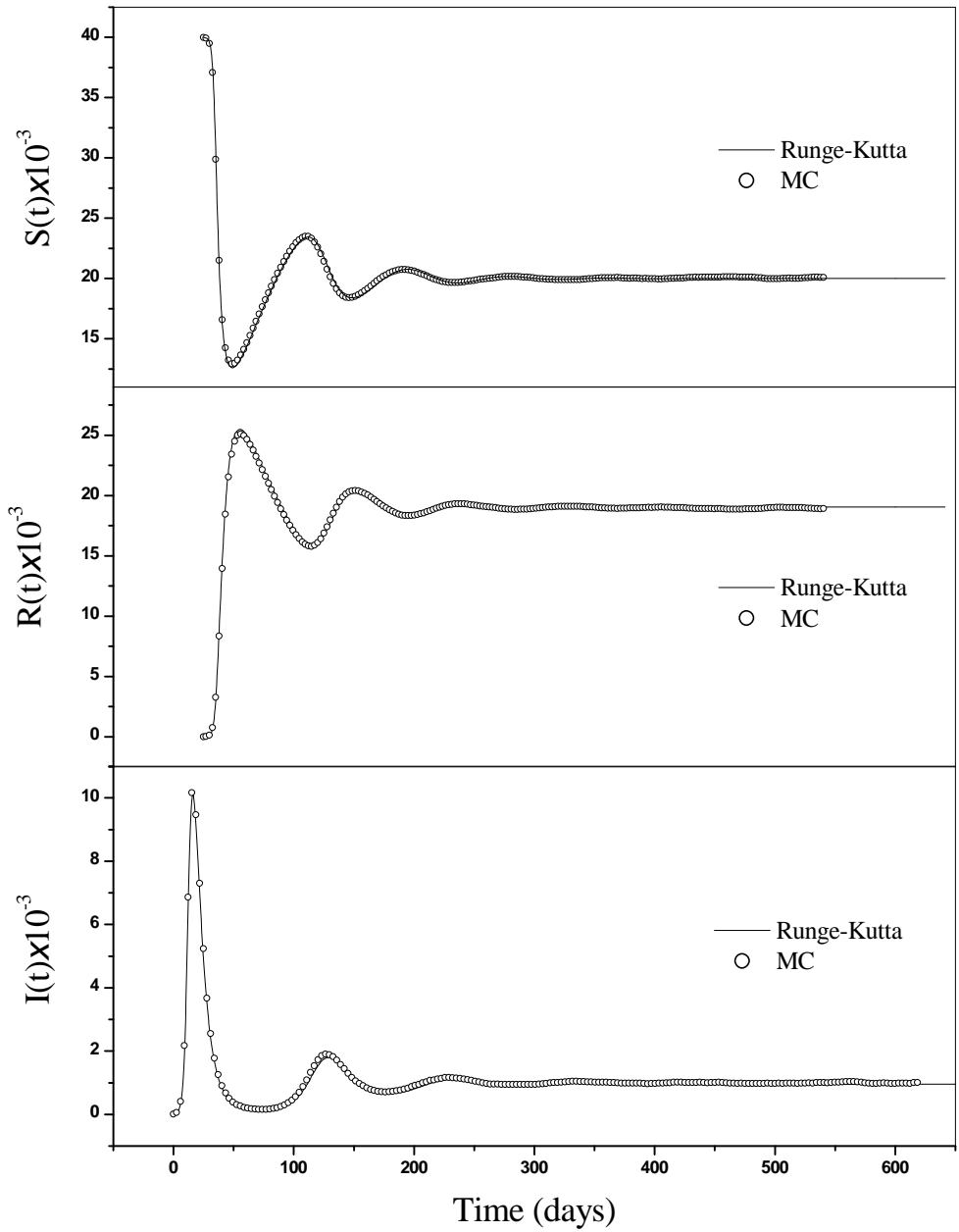
FIG. 2. $SIRS$ model. The figure shows the time evolution of the $S$, $I$, $%
R $. The parameter values are $r=0.2$, $b=0.8$, $m=0.01$ and $\mu =2$. The
error between the $MC$ results (open circles) and Runge-Kutta calculations
are less than $0.1\%$.

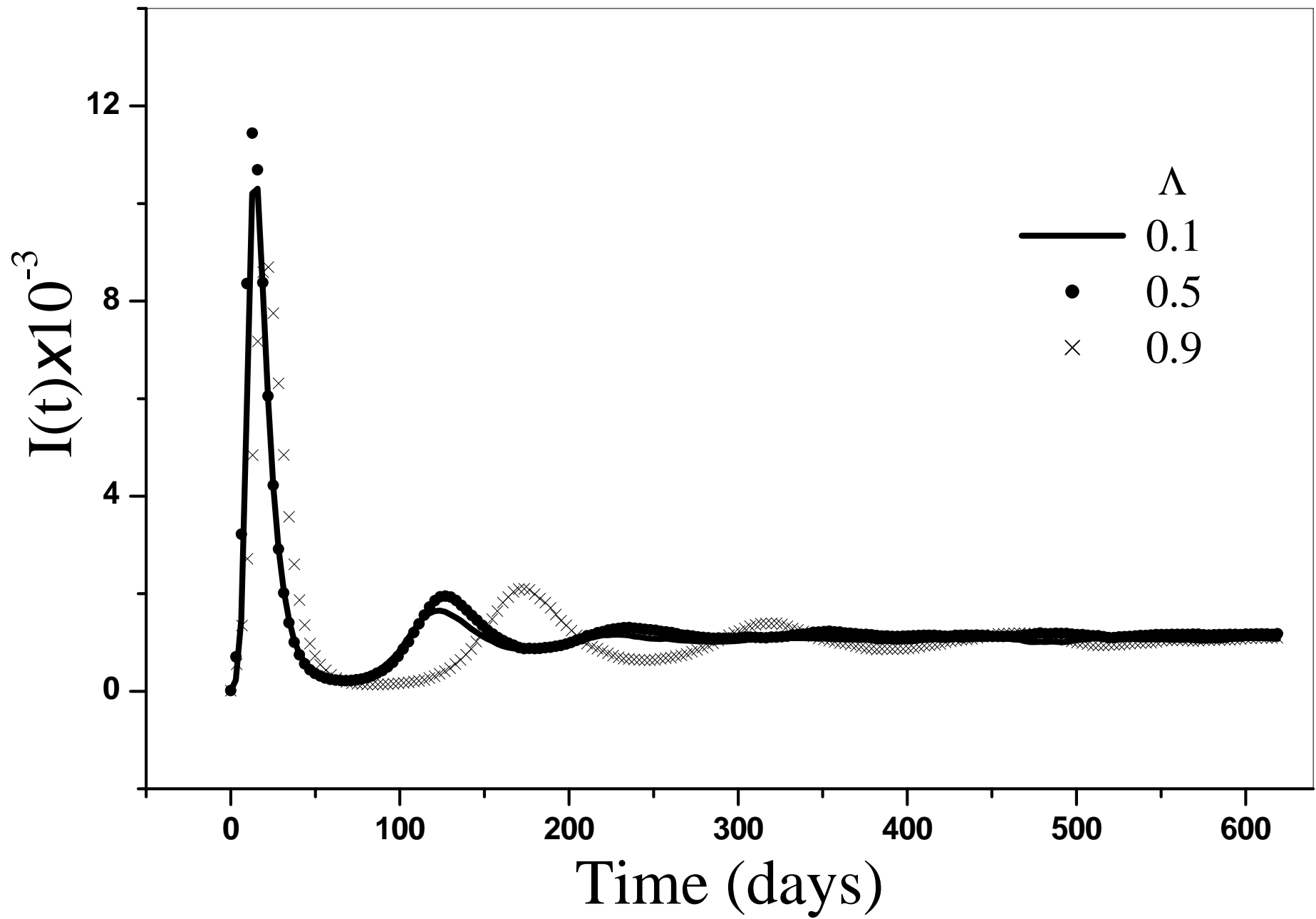
FIG. 3. In this figure, it is shown the effect of spatial variables for the $%
SIRS$ model: the $\Lambda $ and $\Gamma $ parameters balance the local and
global variables $(\Lambda +\Gamma =1)$; the herd immunity effect increases
with $\Lambda $ and it is responsible for the displacement of the curve
second peak to the right.

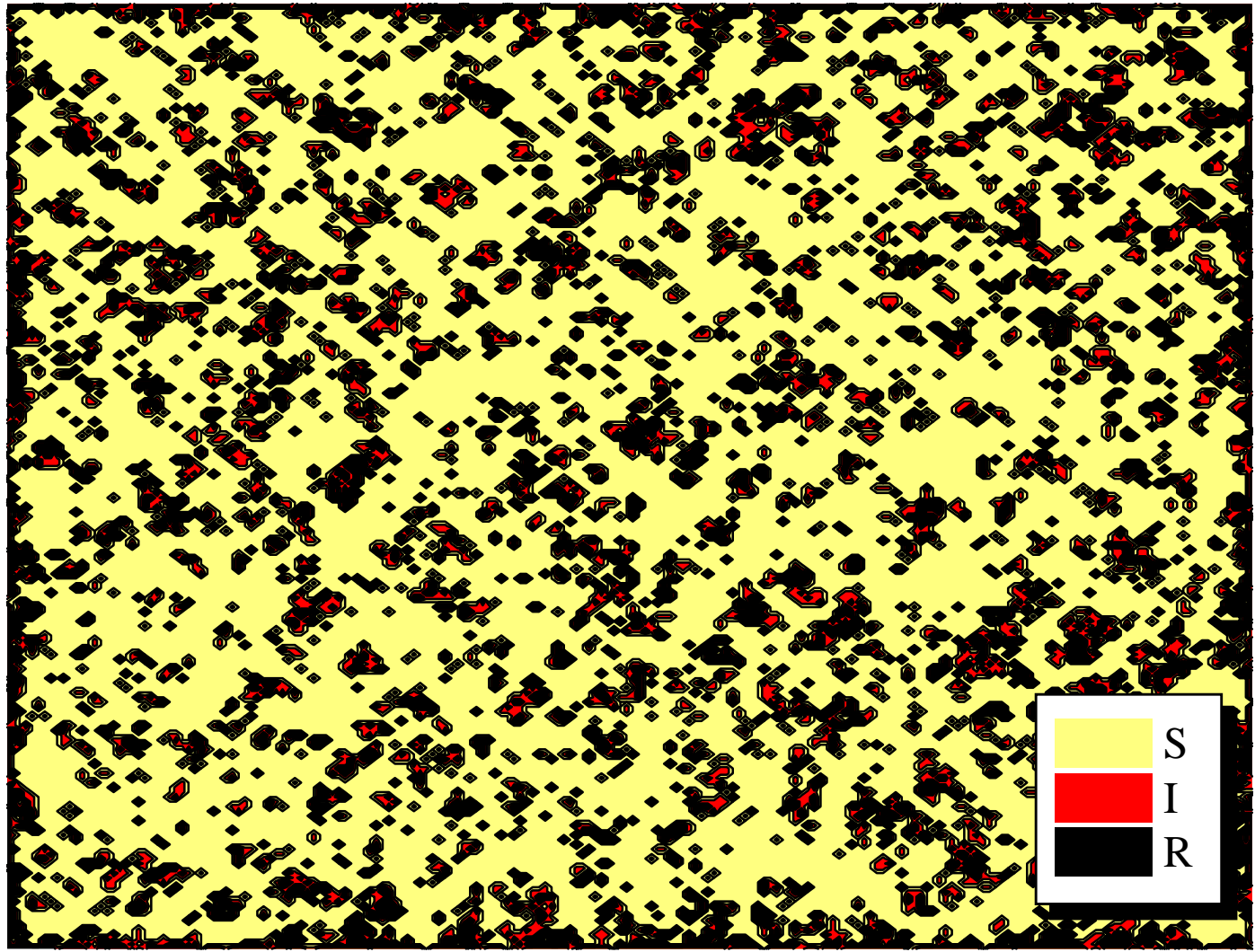
FIG. 4. The shield effect: a snapshot of the system evolution at $t\simeq 40$%
, when $S=19400$ (yellow surface), $R=19200$ (black) and $I=1400$ (red
spots) for $\Lambda =0.9$.


\begin{thebibliography}{99}
\bibitem{footnote} This approach was applied to the Stochastic Euler
procedure too.
\end{thebibliography}
\end{document}